\documentclass[twocolumn,showpacs]{revtex4}
\usepackage{bm,graphicx,amsmath}
\usepackage{epsfig}

\begin{document} 

\title{Jahn-Teller systems from a cavity QED perspective}

\author{Jonas Larson}
\affiliation{NORDITA, 106 91 Stockholm,
Sweden}

\date{\today}

\begin{abstract}
Jahn-Teller systems and the Jahn-Teller effect are discussed in terms of cavity QED models. By expressing the field modes in a quadrature representation, it is shown that certain setups of a two-level system interacting with a bimodal cavity are described by the Jahn-Teller $E\!\times\!\varepsilon$ Hamiltonian. We identify the corresponding adiabatic potential surfaces and the conical intersection. The effects of a non-zero geometrical Berry phase, governed by encircling the conical intersection, are studied in detail both theoretically and numerically. The numerical analysis is carried out by applying a wave packet propagation method, more commonly used in molecul or chemical physics, and analytic expressions for the characteristic time scales are presented. It is found that the collapse-revival structure is greatly influenced by the geometrical phase and as a consequence, the field intensities contain direct information about this phase. We also mention the link between the Jahn-Teller effect and the Dicke phase transition in cavity QED.       
\end{abstract}


\maketitle

\section{Introduction}
The {\it Jahn-Teller} (JT) effect, due to Hermann Jahn and Edward Teller \cite{JTorig}, states that a symmetry-breaking is likely (only exceptions are linear molecules or molecules possessing {\it Kramers degeneracy points} \cite{JTeff2}) to occur whenever there is an isolated degeneracy of electronic states in a molecule, a so called {\it conical intersection} (CI) \cite{ci,cibook}. Over the years, the JT effect has gained enormous attention, mainly in molecular and condensed matter physics \cite{JTeff2,JTeff1,JTcrystal}. A simple model system Hamiltonian possessing a CI, later termed $E\!\times\!\varepsilon$, was presented by H. C. Longuet-Higgens {\it et al.} \cite{longuet-higgens}. The main result of this work states that the angular momentum quantum number is half integer valued rather than an integer. This phenomenon arises from a {\it geometric phase}, on top of the dynamical one, obtained while encircling the CI. The additional phase must be introduced in order to have a single valued total (electronic and vibrational) wave function. This was further analyzed in \cite{mt}, where it was shown that the double valuedness of the electronic wave function can indeed be removed by introducing a ``vector potential" term in the Hamiltonian. For CI models, this resembles the {\it Aharanov-Bohm effect} \cite{ab}, and gave rise to the {\it molecular Aharanov-Bohm effect} and {\it molecular gauge theory} \cite{berrymol1,berrymol2}. A deeper understanding of this phase effect was gained with the seminal paper of M. V. Berry \cite{berry}, which presents a general formalism for the geometrical phase factors that an adiabatic change in the Hamiltonian brings about. In the spring from this work came several papers on the geometrical phase related to CIs, see Refs. \cite{berrymol1,berrymol2}. 

The effect of the geometrical phase on physical observables has since then been discussed and experimentally verified in several reports \cite{faseffects,singlesurf}. The modulation, caused by the geometrical phase, of the wave function has been addressed in Ref. \cite{koppel}, where a dynamical wave packet approach, like the one used in this article, is applied to the $E\!\times\!\varepsilon$ JT model. Inclucion of spin-orbit couplings has been investigated in terms of the JT effect \cite{JTso} and of wave packets \cite{spinorbit2}. Recently, other properties of JT models, not only the $E\!\times\!\varepsilon$, have been considered, for example, quantum chaos \cite{chaos} and ground state entanglement \cite{ent,magnetic2}. 

Although the JT effect has not, to the best of my knowledge, been discussed within {\it cavity quantum electrodynamics} (QED), the geometrical Berry phase has been analyzed in the framework of cavity QED \cite{berrycav,vedralham,berrycav2}. These references study the effects induced by the vacuum field on the geometrical phase. In other words, the treatment of the two-level particle in the time-varying field is then considered on a fully quantum mechanical footing. The degeneracy point is not, however, identified as a CI in these works, and the {\it rotating wave approximation} (RWA) has been applied which is not the case in the present article. In addition, in the situation studied here, the geometrical phase is said to be of {\it dynamical character} as it originates from the intrinsic evolution of the system \cite{berrymol2} rather than from ``external" changing of the Hamiltonian  \cite{berrycav,vedralham,berrycav2}. Thus, the circumstances and approaches are notably different between this work and the ones of Refs. \cite{berrycav,vedralham,berrycav2}. Among others, we examine the cavity QED system in a {\it conjugate representation} in which the intracavity fields are expressed in their quadrature operators rather than the standard used creation and annihilation ladder operators. In this picture, the link to JT systems and to CIs is revealed and, in fact, the {\it Dicke normal-superradiant phase transition} in cavity QED is seen to be related to the JT effect.

The model system is a two-level quantum-dot embedded in a cavity and interacting with two degenerate field modes. The numerical analysis is carried out using wave packet propagation methods; an initial state of the system is let to evolve under the corresponding Hamiltonian. The full Hamiltonian dynamics is considered and compared to results obtained from a second Hamiltonian which shares the same {\it adibatic potential surfaces} (APS) but lacks a geometrical phase. We give analytical expressions for the characteristic time scales, collapse and revivals, for both systems and it is found that the revival time is in a sense twice as long in the case where the geometrical phase is excluded. A consequence of this is reflected in the intracavity field intensities. Thus, measurements of the field intensities of the two modes give a direct indication of the geometrical phase. 

The outline of the paper is as follows. In Sec.~\ref{sec2} we review some basics of JT models and CIs. Especially, in Subsec. \ref{subsec2a} we introduce the $E\!\times\!\varepsilon$ Hamiltonian and discuss its APSs, while Subsec.~\ref{subsec2b} derives the geometrical phase accumulated by encircling the CI, and Subsec.~\ref{subsec2c} considers the JT effect in general and in the $E\!\times\!\varepsilon$ Hamiltonian in particular. The next Sec.~\ref{sec3} is devoted to our cavity QED model and it is shown how the $E\!\times\!\varepsilon$ model occurs for a two-level system interacting with degenerate bimodal fields. A discussion of the corresponding JT effect in cavity QED is outlined in Subsec.~\ref{subsec3b}, where a parallel with the Dicke normal-superradiant phase transition is drawn. Our numerical results of the cavity JT system are presented in Sec.~\ref{sec4}, both analyzing the short and long term behavior and how the geometrical phase comes into play. Finally we summarize in Sec.~\ref{sec5}.

\section{The Jahn-Teller model}\label{sec2}
Jahn-Teller systems are characterized by a degeneracy point of
coupled potential surfaces, a CI.
In one dimension, the simplest example is the $E\!\times\!\beta$
model, also called {\it Rabi} or {\it spin-boson} model
\cite{oldwine}. It describs a spin 1/2 particle coupled to a single
boson mode \cite{eb}. In certain parameter regimes, a RWA can be applied in which this model relaxes to the
one of Jaynes and Cummings \cite{jc}. In one dimension, the wave packet (state of
the system) cannot encircle the CI without passing through it, and therefore there is no corresponding
dynamical geometric phase \cite{berry}. In two dimensions, generalization of the
$E\!\times\!\beta$ model leads in certain situations to the $E\!\times\!\varepsilon$ model which will be
the subject of this section.

\subsection{The $E\!\times\!\varepsilon$ Hamiltonian}\label{subsec2a}
The simplest Jahn-Teller Hamiltonian with two vibrational degrees of freedom is the so called $E\!\times\!\varepsilon$
one, given by \cite{longuet-higgens}
\begin{equation}\label{jtham}
\begin{array}{lll}
H_{JT} & = &
\displaystyle{\frac{\hat{p}_x^2}{2m}+\frac{\hat{p}_y^2}{2m}+\frac{m\omega^2}{2}(\hat{x}^2+\hat{y}^2)}
\\ \\
& &
\displaystyle{+\lambda\hat{x}\hat{\sigma}_x+\lambda\hat{y}\hat{\sigma}_y}.
\end{array}
\end{equation}
Here $\hat{p}_i$ and $\hat{x}_i$ are momentum and position in the
$i$ direction of the ``particle" with mass $m$. The
$\hat{\sigma}$-operators are the standard Pauli matrices obeying the
commutation relations
\begin{equation}
[\hat{\sigma}_i,\hat{\sigma}_j]=2\varepsilon_{ijk}\hat{\sigma}_k
\end{equation}
and with the $z$-eigenstates,
$\hat{\sigma}_z|\pm\rangle=\pm|\pm\rangle$ and $\lambda$ the coupling constant. Clearly, at the origin
$\hat{x}=\hat{y}=0$, the two potential surfaces are degenerate. In
the presence of either spin-orbit coupling
\cite{spinorbit1,spinorbit2} or an external magnetic field
\cite{magnetic1,magnetic2}, an additional {\it detuning} term
$\Delta\hat{\sigma}_z/2$ is added to the Hamiltonian, where $\Delta$
is the spin-orbit splitting or the magnetic strength. With this term present, the degeneracy is lifted and the intersection becomes avoided. 

The form of the Hamiltonian (\ref{jtham}) defines the {\it diabatic
basis} and {\it diabatic potentials}, namely; a diabatic state is
written as $\Psi(x,y)=f_\pm(x,y)|\pm\rangle$ for some normalized
function $f_\pm(x,y)$, and the diabatic potentials, once the detuning $\Delta$ is included, are
$m\omega^2(x^2+y^2)/2\pm\Delta/2$. Before defining the APSs, we express the Hamiltonian in polar
coordinates
\begin{equation}
x\pm iy=\rho\mathrm{e}^{\pm i\varphi}
\end{equation}
giving \cite{koppel,spinorbit2,spinorbit1} 
\begin{equation}\label{jtpolar}
\begin{array}{lll}
H_{JT} & = &
\displaystyle{-\frac{\hbar^2}{2m}\left(\frac{\partial^2}{\partial\hat{\rho}^2}+
\frac{1}{\hat{\rho}}\frac{\partial}{\partial\hat{\rho}}+
\frac{1}{\hat{\rho}^2}\frac{\partial^2}{\partial\hat{\varphi}^2}\right)}
\\ \\
& &
\displaystyle{+\frac{m\omega^2}{2}\hat{\rho}^2+\left[\begin{array}{cc}\displaystyle{\frac{\Delta}{2}}
& \lambda\hat{\rho}\mathrm{e}^{i\hat{\varphi}} \\
\lambda\hat{\rho}\mathrm{e}^{-i\hat{\varphi}} &
-\displaystyle{\frac{\Delta}{2}}\end{array}\right]}
\end{array}
\end{equation}

\begin{figure}[ht]
\begin{center}
\includegraphics[width=8cm]{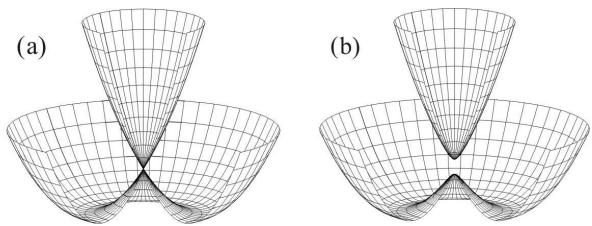}
\vspace{0cm} \caption{Two examples of the adiabatic surfaces in the
$E\!\times\!\varepsilon$ Jahn-Teller model with detuning $\Delta=0$
(a) and $\Delta\neq0$ (b).  } \label{fig1}
\end{center}
\end{figure}

Let us introduce the unitary operator \cite{koppel,mt}
\begin{equation}\label{adstate}
U=\left[\begin{array}{cc} \sin(\nu)
& \cos(\nu) \\
-\cos(\nu)\mathrm{e}^{i\mu} & \sin(\nu)\mathrm{e}^{i\mu}
\end{array}\right],
\end{equation}
where
\begin{equation}
\tan(2\nu)=\frac{2\lambda\hat{\rho}}{\Delta}
\end{equation}
and $\mu=\hat{\varphi}$, which diagonalizes the last term of Eq.
(\ref{jtpolar}). However, $U$ does not commute with the kinetic
term in Eq.~(\ref{jtpolar}) and consequently, the transformed
Hamiltonian, $\tilde{H}_{JT}=U^{-1}H_{JT}U$, is non-diagonal. The
off-diagonal terms are the non-adiabatic couplings, which usually
are small far from the crossing. Omitting these terms defines the 
{\it adiabatic Hamiltonian}
\begin{equation}\label{jtpolarad}
H_{JT}^{ad}=T+V_\pm^{ad}+V_{cent}+V_{gauge},
\end{equation}
where \cite{koppel,mt}
\begin{equation}\label{jtpolaradterms}
\begin{array}{l}
T=\displaystyle{-\frac{\hbar^2}{2m}\left(\frac{\partial^2}{\partial\hat{\rho}^2}+
\frac{1}{\hat{\rho}}\frac{\partial}{\partial\hat{\rho}}+\frac{1}{\hat{\rho}^2}\frac{\partial^2}{\partial\hat{\varphi}^2}\right),}
\\ \\
V_\pm^{ad}=\displaystyle{+\frac{m\omega^2}{2}\hat{\rho}^2+\hat{\sigma}_z\sqrt{\left(\!\frac{\Delta}{2}\right)^2+\lambda^2\hat{\rho}^2},}
\\ \\ V_{cent}=\displaystyle{\frac{\lambda^2\Delta^2\omega}{2\left(\Delta^2+4\lambda^2\hat{\rho}^2\right)^2},}
\\ \\
V_{gauge}=\displaystyle{\frac{\omega}{2\hat{\rho}^2}\left(1+\frac{\Delta}{\sqrt{\Delta^2+4\lambda^2\hat{\rho}^2}}\right)\left(\frac{1}{2}-\frac{\partial}{\partial\hat{\varphi}}\right)}.
\end{array}
\end{equation}
We will assume that the evolution takes place mainly upon the lower
APS, {\it i.e.}, $\hat{\sigma}_z=-1$, but in the
numerical simulations we consider full dynamics without
any approximations. Close to the crossing, the non-adiabatic couplings may have a significant impact on the dynamics \cite{cibook,bobreak}. The term $V_\pm^{ad}$ defines the APSs, while $V_{cent}$ and $V_{gauge}$, arising from the commutator between $U$ and the kinetic
energy operator, are {\it centrifugal corrections} \cite{centcorr}. The last term has been labeled {\it gauge} for reasons that will become clear later on.
We display two examples, $\Delta=0$ and $\Delta\neq0$, of the
APSs $V_\pm^{ad}$ in Fig.~\ref{fig1}. For a non-zero detuning, as pointed out,
the crossing at the CI becomes avoided, with
splitting amplitude $\Delta$. The lower surface has the familiar sombrero shape, while the upper possesses a single global minimum at
$x=y=0$. For large detunings $\Delta$, the Mexican hat structure is
lost, and the minimum of the lower APS is at the origin. Especially, the
radius giving the potential minima is given by \cite{spinorbit2}
\begin{equation}\label{potmin}
\rho_{min}=\left\{\begin{array}{lll}
\displaystyle{\sqrt{\left(\frac{\lambda}{\omega}\right)^2-\left(\frac{\Delta}{\lambda}\right)^2}},
& \hspace{1cm} & \omega|\Delta|<\lambda^2 \\ \\
0, & \hspace{1cm} & \omega|\Delta|\ge\lambda^2
\end{array}\right..
\end{equation}

\subsection{Berrys geometrical phase in the
$E\!\times\!\varepsilon$ model}\label{subsec2b}
The {\it adiabatic states}, defined
by $U$, are arbitrary up to an overall phase. The phase choice in Eq.
(\ref{adstate}) is chosen such that the states are singled valued as
$\varphi$ is varied by $2\pi$. For example, the alternative obtained by multiplying $U$ by
$\exp\left(-i\varphi/2\right)$ implies double-valued adiabatic
states. Unitary transformation of the Hamiltonian in this second case leads
to an adiabatic Hamiltonian lacking the term proportional to
$i\partial/\partial\hat{\varphi}$. The very last term of
(\ref{jtpolarad}), containing the differential operator
$\partial/\partial\hat{\varphi}$, can be viewed as a vector
potential. Indeed, this term can be combined with the canonical momenta to define a kinetic momenta. Thus, the two options of overall phase of the adiabatic
states given above result in either single-valued states with a vector
potential present in the Hamiltonian or no such vector potential term but
double-valued states \cite{vector}. The source of a vector potential term is the cause for having a non-zero geometric phase as the system encircles the CI in analogous to the Aharanov-Bohm effect \cite{spinorbit2}. 

For the system evolving along a closed loop $C$ in parameter space, the
geometrical phase can be calculated according to \cite{berry}
\begin{equation}
\gamma_n(C)=\oint_C\langle n({\bf R})|\nabla_{{\bf R}}n({\bf
R})\rangle\cdot d{\bf R}.
\end{equation}
Here, $|n({\bf R})\rangle$ is the $n$'th adiabatic eigenstate and ${\bf
R}$ the set of parameters. In particular, in our case $|n({\bf
R})\rangle=\left(\sin(\theta),-\cos(\theta)\mathrm{e}^{i\varphi}\right)$
and as we consider a time-independent dynamical problem, the varying
parameters ${\bf R}$ are the coordinates $\rho$ and $\varphi$.
Especially, we consider a wave packet located at the minima of the
sombrero potential, such that
$\langle\hat{\rho}\rangle\approx\rho_{min}$, and $\varphi$ is changed from
0 to $2\pi$. For a general radius $R$ we find the
$E\!\times\!\varepsilon$ geometric phase \cite{spinorbit2}
\begin{equation}
\gamma_{JT}(R)=-\pi\left(1+\frac{\Delta}{\sqrt{\Delta^2+\lambda^2R^2}}\right),
\end{equation}
which for $R=\rho_{min}$ becomes
\begin{equation}\label{berryfas}
\gamma_{JT}(\rho_{min})=-\pi\left(1+\frac{\omega\Delta}{\lambda^2}\right).
\end{equation}
For $\Delta=0$ we obtain the well known sign change of the wave
function when encircling a CI, causing half
integer angular momentum quantum numbers.

\subsection{The Jahn-Teller effect}\label{subsec2c}
Using group theoretical arguments, Jahn and Teller proved that for almost any degeneracy (CIs) among electronic states in a molecule, a symmetry-breaking is ``allowed" which removes the degeneracy and lowers the total energy of the model system ground state \cite{JTorig}. It turns out that this symmetry-breaking indeed takes place in the majority of cases with some exceptions \cite{JTeff2,JTeff1}. Hence, the molecule favors a distortion of its most symmetrical state. This effect is quenched when spin-orbit coupling is taken into account, but may still exist \cite{JTso}. Returning to the APSs of the $E\!\times\!\varepsilon$ Hamiltonian (\ref{jtpolarad}), the state with highest symmetrical is a wave packet centered around the origin ($x=y=0$). Loosely speaking, for vanishing detuning $\Delta$, is it intuitive to expect the wave packet to slide down the potential surfaces towards the minima of the sombrero. Semi-classically, the wave packet experiences a non-zero force $F=-\nabla V(x,y)$. For large detunings however, we saw from Eq. (\ref{potmin}) that the potential surfaces may possess a single global minimum which will prevent the JT-distortion. On the other hand, for small but non-zero detuning, the sombrero structure is present for the lower APS and here quantum fluctuations will permit a symmetry breaking. Naturally, the above arguments are semi-classical and the full evolution is quantum mechanical and described by the coupled system. Nonetheless, it gives some insight and intuition of the dynamics.

\section{Jahn-Teller models in cavity QED}\label{sec3}

Spin boson models naturally occur in cavity QED. Here the boson
subsystem represents a single or several quantized modes of an intracavity field, while the spin
degrees of freedom describes either two-level atom/atoms \cite{haroche}
or solid state quantum-dot/dots \cite{qdot}. Contrary to standard
formulations of Jahn-Teller models, here the bosons are the photons
of the field rather than vibrational phonons, and the internal
structure corresponds to two discrete energy levels of the atom or
quantum-dot. In the single mode case, a microscopic derivation gives the
$E\!\times\!\beta$ Hamiltonian \cite{scully} in the assumption of
{\it dipole approximation} and neglecting the {\it self-energy} (see
below). In most cavity QED experiments involving atoms, the
application of the RWA is
justified, in which the Hamiltonian identifies the analytically
solvable Jaynes-Cummings one \cite{jc}. The APSs (or rather adiabatic potential curves) of the Jaynes-Cummings model contain the
differential momentum operator \cite{oldwine}, and they are therefore said to be of {\it non-potential form}. Nonetheless, even if the picture of potential surfaces is less intuitive due to the momentum dependence, the JC model renders a sort of generalized CI (curve crossing). To go beyond
the RWA regime, the coupling to the field must be substantially increased
compared to atomic cavity QED setups. This is indeed the case for
solid state quantum-dots coupled to a cavity. In fact, the crucial
parameter, coupling divided by the two-level transition frequency,
can be made several orders of magnitude larger in condensed matter
systems \cite{qdot} compared to atom-cavity ones. Another possibility to achieve ultrastrong atom-field couplings is to consider Bose-Einstein condensates coupled to an intracavity field \cite{bec}.  

\subsection{The model system Hamiltonian}\label{subsec3a}

\begin{figure}[ht]
\begin{center}
\includegraphics[width=8cm]{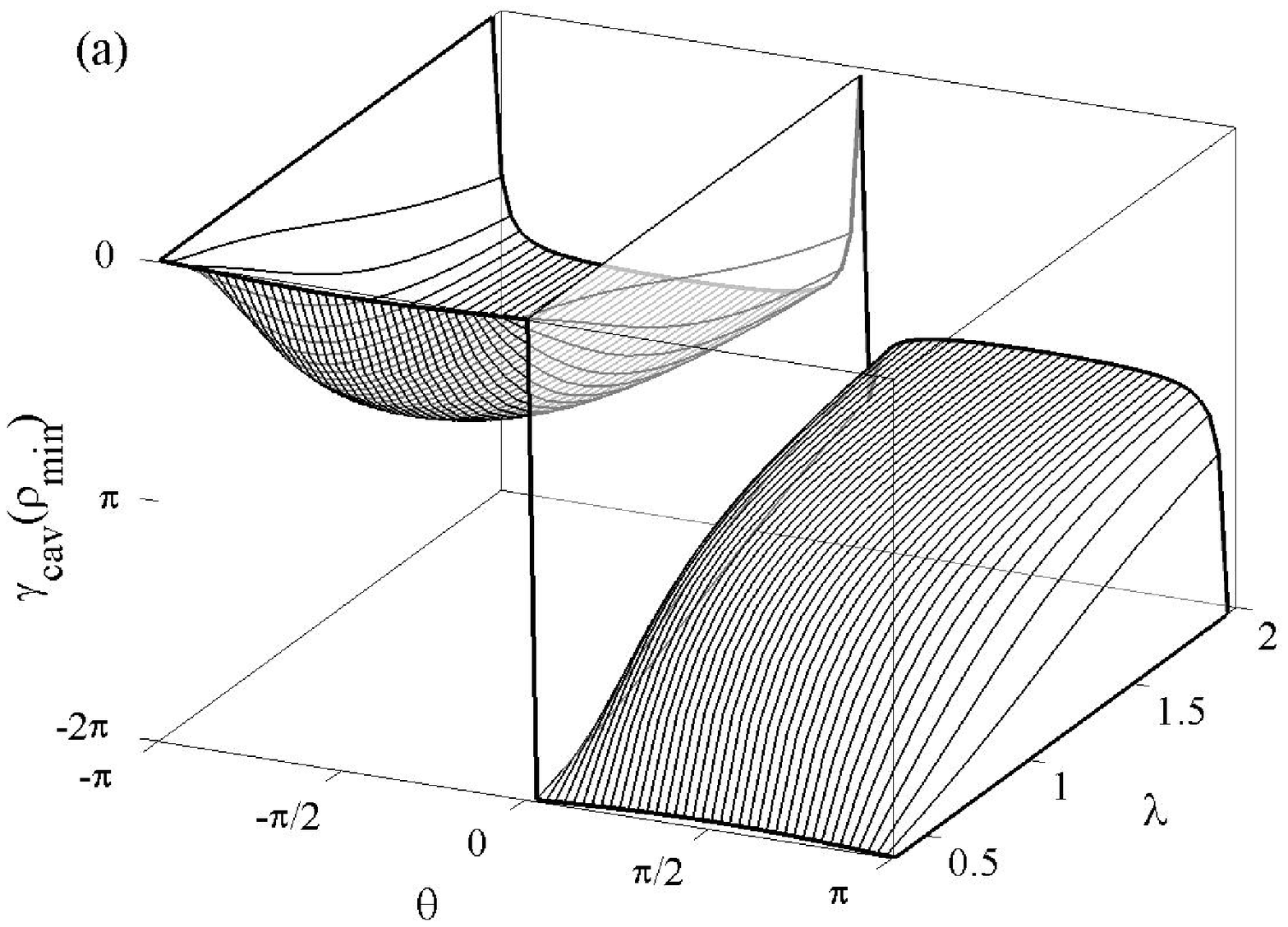}
\includegraphics[width=8cm]{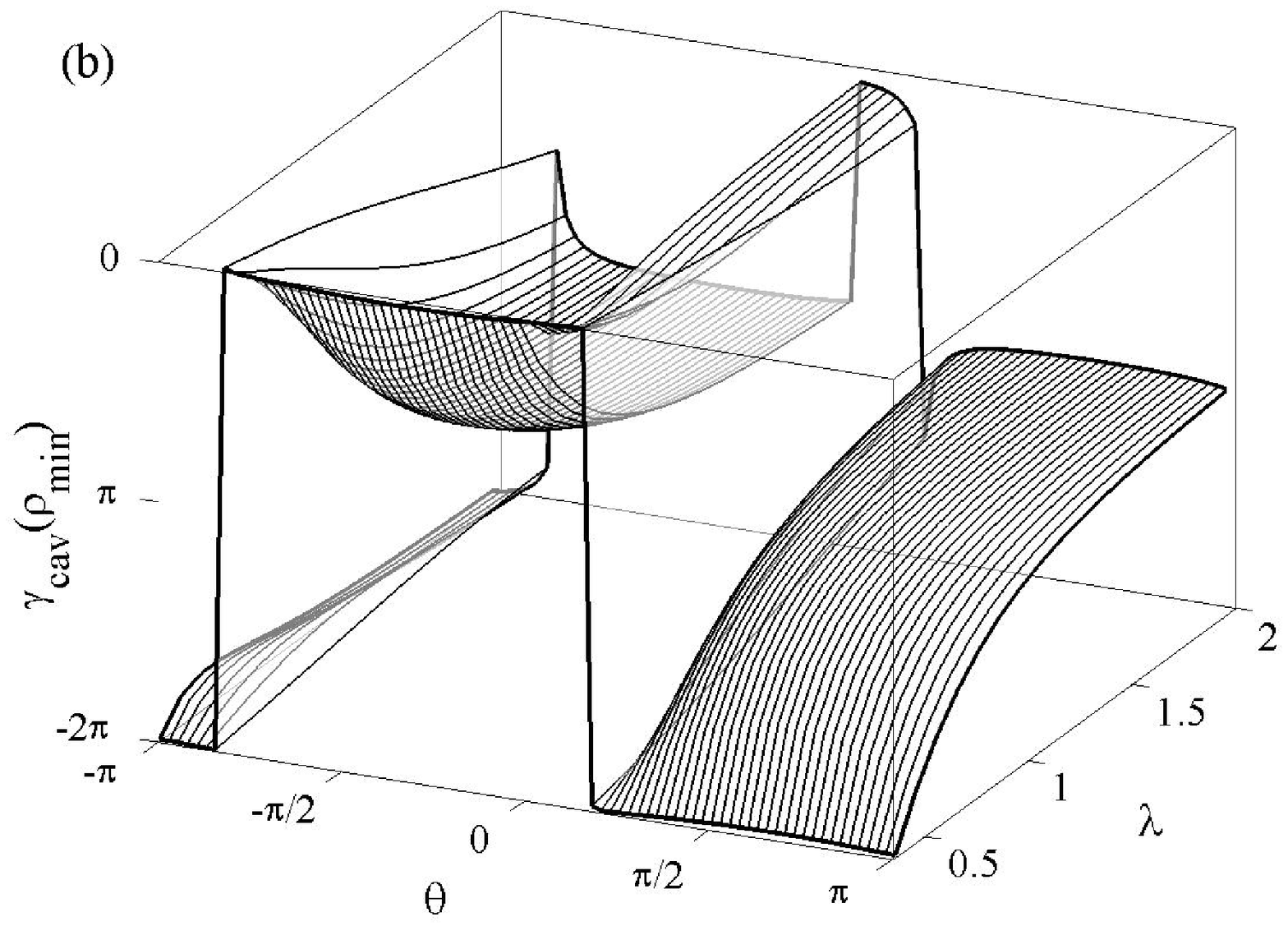}
\vspace{0cm} \caption{The geometric phase (\ref{berryfas2}) as
function of coupling $\lambda$ and field phase $\theta$. The radius
$R=\rho_{min}$, where
$\rho_{min}=\sqrt{4\lambda^2-(\Omega/4\lambda)^2}$ is the adiabatic
potential minima of the sombrero when $\theta-\phi=\pi/2$ (the
cylindrically symmetric case). The other dimensionless parameters
are $\phi=0$ and $\Omega=1$ (a) and $\phi=1/2$ and $\Omega=1$ (b). }
\label{fig2}
\end{center}
\end{figure}

To obtain CIs rather than curve crossings, multi-mode
cavities must be considered \cite{twomode}. For simplicity we will
assume two degenerate cavity modes such that they share the same frequency
$\tilde{\omega}$ and also same coupling amplitude $\tilde{\lambda}$ to
the quantum-dot. The Hamiltonian in the dipole approximation reads
\cite{hamref,hamref2}
\begin{equation}\label{hamcav1}
\begin{array}{lll}
H_{cav} & = &
\displaystyle{\hbar\tilde{\omega}\left(\hat{a}^\dagger\hat{a}+\hat{b}^\dagger\hat{b}\right)+\hbar\frac{\tilde{\Omega}}{2}\hat{\sigma}_z}\\
\\
& & \displaystyle{+\hbar
\frac{\tilde{\lambda}}{\sqrt{2}}\Big[\left(\hat{a}^\dagger+\hat{a}\right)\left(\hat{\sigma}^+\mathrm{e}^{-i\phi}+\hat{\sigma}^-\mathrm{e}^{i\phi}\right)}\\
\\
& & \displaystyle{+
\left(\hat{b}^\dagger+\hat{b}\right)\left(\hat{\sigma}^+\mathrm{e}^{-i\theta}+\hat{\sigma}^-\mathrm{e}^{i\theta}\right)\Big]}.
\end{array}
\end{equation}
Here $\hat{a}^\dagger$ and $\hat{b}^\dagger$ ($\hat{a}$ and $\hat{b}$)
are creation (annihilation) operators for the two field modes,
$\phi$ and $\theta$ field phases, $\tilde{\Omega}$ the quantum-dot
transition frequency, and $2\hat{\sigma}^\pm=\hat{\sigma}_x\pm
i\hat{\sigma}_y$. In the following we will label the two cavity modes by $a$ and $b$. Before proceeding, for brevity we introduce a
characteristic energy $\hbar\tilde{\omega}$ and time scale
$\tilde{\omega}^{-1}$, such that we consider dimensionless variables
\begin{equation}
\begin{array}{lllll}
\displaystyle{\lambda=\frac{\tilde{\lambda}}{\tilde{\omega}}}, & &
\displaystyle{\Omega=\frac{\tilde{\Omega}}{\tilde{\omega}}}, & &
\tau=\tilde{\omega}t,
\end{array}
\end{equation}
where $t$ is the unscaled time. In a {\it conjugate variable
representation} defined by the operator relations
\begin{equation}\label{convar}
\begin{array}{c}
\displaystyle{\hat{p}_x=i\frac{1}{\sqrt{2}}\left(\hat{a}^\dagger-\hat{a}\right)},\hspace{1cm}
\displaystyle{\hat{x}=\frac{1}{\sqrt{2}}\left(\hat{a}^\dagger+\hat{a}\right)},\\
\\
\displaystyle{\hat{p}_y=i\frac{1}{\sqrt{2}}\left(\hat{b}^\dagger-\hat{b}\right)},\hspace{1cm}
\displaystyle{\hat{y}=\frac{1}{\sqrt{2}}\left(\hat{b}^\dagger+\hat{b}\right)},
\end{array}
\end{equation}
where $[\hat{x},\hat{p}_x]=[\hat{y},\hat{p}_y]=i$, the Hamiltonian
(\ref{hamcav1}) takes the form
\begin{equation}\label{hamcav2}
\begin{array}{lll}
H_{cav} & = &
\displaystyle{\frac{\hat{p}_x^2}{2}+\frac{\hat{p}_y^2}{2}+\frac{\hat{x}^2}{2}+\frac{\hat{y}^2}{2}+\frac{\Omega}{2}\hat{\sigma}_z}
\\ \\
& &
\displaystyle{+2\lambda\hat{x}\big[\cos(\phi)\hat{\sigma}_x+\sin(\phi)\hat{\sigma}_y\big]}
\\ \\
& &
\displaystyle{+2\lambda\hat{y}\big[\cos(\theta)\hat{\sigma}_x+\sin(\theta)\hat{\sigma}_y\big]}.
\end{array}
\end{equation}
For the simple example of $\phi=0$ and $\theta=\pi/2$ we recover the cylindically symmetric
$E\!\times\!\varepsilon$ Hamiltonian (\ref{jtham}). In fact, for
\begin{equation}\label{eecon}
|\phi-\theta|=(j+1/2)\pi,\hspace{1.2cm} j
\hspace{0.3cm}\mathrm{integer},
\end{equation}
$H_{cav}$ is unitarilly equivalent with the $E\!\times\!\varepsilon$ Hamiltonian $H_{JT}$ by identifying $\Omega$ with $\Delta$. In some special cases of the phases, the last two terms of (\ref{hamcav2}) can be written as 
\begin{equation}
\begin{array}{l}
2\lambda(\hat{x}+\hat{y})\left[\begin{array}{cc}
0 & \mathrm{e}^{-i\phi}\\
\mathrm{e}^{i\phi} & 0\end{array}\right],\hspace{0.7cm}\mathrm{for}\hspace{0.2cm}\theta-\phi=2j\pi,\\ \\
2\lambda(\hat{x}-\hat{y})\left[\begin{array}{cc}
0 & \mathrm{e}^{-i\phi}\\
\mathrm{e}^{i\phi} & 0\end{array}\right],\hspace{0.7cm}\mathrm{for}\hspace{0.2cm}\theta-\phi=(2j+1)\pi,
\end{array}
\end{equation}
for some integer $j$. For these situations the CI is replaced by an intersecting curve in the directions of $\varphi=3\pi/4,\,7\pi/4$ or $\varphi=\pi/4,\,5\pi/4$ respectively. Here, it is clear that by a unitary rotation, the adiabatic states can be made real, indicating that the geometrical phase becomes identically zero as the wave packet is encircling the CI. This is indeed seen in Fig.~\ref{fig2} displaying the geometric phases of the Hamiltonian (\ref{hamcav2}). The general form of the APSs, in polar coordinates, reads
\begin{equation}
V_\pm^{ad}(\rho,\varphi)=\!\frac{\rho^2}{2}\!\pm\!\sqrt{\!\left(\frac{\Omega}{2}\right)^{\!2}\!\!+4\lambda^2\rho^2\!\left[1\!+\!\cos(\phi\!-\!\theta)\sin(2\varphi)\right]}.
\end{equation}
The lower APS has two minima for angels
$\varphi=\pi/4,\,3\pi/4$. It is known that the ``quadratic" $E\!\times\!\varepsilon$
Hamiltonian has three local minima in the sombrero shaped potential \cite{koppel,spinorbit2}. This derives from a term of the form $\sin(3\varphi)$ in the APSs. Here we have a $\sin(2\varphi)$-dependence instead and hence the double minima structure. 

The single valued adiabatic states can again be written 
\begin{equation}\label{adstate2}
\Psi_u(\rho,\varphi)=\!\left[\!\begin{array}{c}\sin(\nu) \\ -\cos(\nu)\mathrm{e}^{i\mu}\end{array}\!\right]\!,\hspace{0.5cm}\Psi_l(\rho,\varphi)=\!\left[\!\begin{array}{c}\cos(\nu) \\ \sin(\nu)\mathrm{e}^{i\mu}\end{array}\!\right],
\end{equation}
but with
\begin{equation}
\begin{array}{l}
\displaystyle{\tan(2\nu)=\frac{4\lambda\rho}{\Omega}\sqrt{1+\sin(2\varphi)\cos(\phi-\theta)}},\\
\\
\displaystyle{\mu=\tan\left(\frac{\cos(\varphi)\sin(\phi)+\sin(\varphi)\sin(\theta)}{\cos(\varphi)\cos(\phi)+\sin(\varphi)\cos(\theta)}\right)}.
\end{array}
\end{equation}
Encircling the CI at a radius $R$ we find the
geometric phase (\ref{berryfas})
\begin{equation}\label{berryfas2}
\gamma_{cav}(R)=-\left.\int_0^{2\pi}\cos^2(\alpha)\frac{\partial\psi}{\partial\varphi}\,d\varphi\right|_{\rho=R}.
\end{equation}
The phase (\ref{berryfas2}) is depicted in Fig.~\ref{fig2} as a
function of $\theta$ and $\lambda$ for fixed $\phi=0$ and $\Omega=1$
(a) and $\phi=1/2$ and $\Omega=1$ (b). The asymptotic value for
large couplings $\lambda$ is either $-\pi$ or $0$ (modulo $2\pi$). The radius $R$ is
taken to be the minimum of the cylindrically symmetric case
($\theta=\pi/2$). For $\phi=0$, the greatest effect generated by the
geometrical phase is seen to be in the symmetric case of
$\theta=\pi/2$, while for $\phi\neq0$ the situation becomes more complex. Note that, according to Eq.~(\ref{potmin}), and identifying $\Omega$ with $\Delta$, $\lambda\neq0$ in order to have a sombrero structure of the lower APS, which is the reason why $\lambda$ is not approaching 0. 

\subsection{The Jahn-Teller effect in cavity QED}\label{subsec3b}

The Jahn-Teller effect states that, in the presence of a CI,
lowering the symmetry may be energetically favorable in various
systems, typically for molecules and crystals. In terms of the
$E\!\times\!\varepsilon$ Hamiltonian this is intuitive, since a
wave packet centered at the origin (the CI) will
have a larger ``potential energy" than a wave packet located at
$\rho_{min}$. The corresponding symmetry breaking in cavity QED
implies that the system ground state consists of non-zero fields in
the two modes. At the same time, the quantum-dot is not
entirely in its lower state but in a superposition of its two internal states. This is related to a well
known phenomenon in quantum optics, namely {\it superradiance}
\cite{dicke}. This, of course, comes about due to the strong interaction
between the quantum-dot and the cavity fields. For multi quantum-dot systems and
in the thermodynamic limit, where the number of two-level quantum-dots
and the volume tend to infinity while the density is kept fixed,
this results in a second order quantum phase transition between a {\it
normal} (the field in its vacuum) and a {\it superradiant phase} (a
macroscopic non-zero field) \cite{dicke2}. The critical coupling of this phase transition is given by $\lambda_c=\sqrt{|\Delta|\omega}$ \cite{dicke2}, which indeed follows from Eq.~(\ref{potmin}).

However, it can be shown that for a quantum-dot in which the lower state
is its ground state, the normal-superradiant phase transition is an
artifact from neglecting the self-energy term from the Hamiltonian
\cite{a2}. For a single mode, and in unscaled units, this term is
given by
\begin{equation}
H_{se}=\frac{e^2}{2m}\frac{\pi\hbar}{V\omega}\hat{x}^2,
\end{equation}
where $e$ is the electron charge, $m$ its mass and $V$ the effective
mode volume. This should be compared with the matter-field coupling
\begin{equation}
\lambda=\Omega d\sqrt{\frac{2\pi\hbar}{V\omega}},
\end{equation}
where $d$ is the dipole moment of the transition of
interest in the quantum-dot. It is clear that the self-energy term $H_{se}$ tends to quench
the Mexican hat structure, and further that $\lambda$ and $H_{sc}$
are not fully independent. Indeed, in Ref.~\cite{a2} it is
demonstrated, either using the Thomas-Reiche-Kuhn sum rule or
simple thermodynamical and gauge invariance arguments, that the
sombrero shape cannot be obtained for any set of physical parameters. 

To circumvent this obstacle one may use a two-photon Raman type of
interaction, where three levels of the quantum-dot are coupled through the
cavity mode and an external classical laser field
\cite{ramandicke,jonasdicke}. In the large detuning limit of one of
the internal levels it can be adiabatically eliminated \cite{adbose} and one
arrives at an effective model very similar to the one above. In such procedure one introduces an additional
independent parameter, the detuning $\delta$ of the eliminated
level \cite{cond}. As the detuning enters in the effective matter-field
coupling parameter, but not in the self energy term, these two become
in principle independent, and in particular $\lambda$ can be made large in comparison with $H_{se}$. The effective model, once the detuned
level has been eliminated, contains some {\it Stark shift} terms
that will modify the potential surfaces, but the CI and sombrero
structure are still present. The external laser fields have
the advantage of being easily controllable in terms of system
parameters such as amplitude and phase. A system Hamiltonian
suitable for realizing the Jahn-Teller model can be found in Ref. 
\cite{vedralham}. It should be noted though, that the effective Hamiltonian is in general time dependent, which is prevailed by imposing a RWA. This, however,
induces a ``momentum-dependent" potential surface, but nonetheless, the JT symmetry
breaking is still present in this approximation \cite{dicke2}.

Another possibility to surmount the problem with the self-energy, which is assumed in this paper, is to use the fact that the two internal levels that couple to the
cavity modes are normally highly excited meta stable (Rydberg)
states. For these states, neither the Thomas-Reiche-Kuhn sum rule nor the
thermodynamical arguments apply, and the symmetry breaking
may still occur. We therefore discard the self-energy terms as they
would only modify the frequencies of the harmonic potentials. In general, also for the Raman coupled model, the states involved are highly excited metastable states and the sum rule cannot be applied in those cases either. A benefit of the Raman model, compared to a one photon model, is the higher controllability of the system parameters; especially the diagonal element is a detuning parameter (and not a transition frequency $\Omega$) that can be made small compared to the matter-field coupling. The drawback of an effective Raman model is that the analysis is considerably less intuitive due to the RWA. Here, we therefore choose the simpler model of the two as the physical phenomena may be more easily extracted from it.

\section{Numerical results}\label{sec4}
Contrary to molecular or solid state systems, properties of the cavity fields are in comparison easily measured, for example, phoson distribution \cite{photonmeas}, field quadratures \cite{squezmeas} and, in fact, the whole phase space distribution using quantum tomography \cite{tomo}. Using wave packet propagation methods, the dynamics of such quantities will be studied in this section with emphasizes on the effects emerging from the geometrical phase. As an initial state we take a disentangled one, given in cartesian coordinates by 
\begin{equation}\label{initialstate2}
\Psi(x,y,0)=\psi(x,y,0)\frac{1}{\sqrt{2}}\left[\begin{array}{c} 1 \\ -1
\end{array}\right],
\end{equation}
where 
\begin{equation}\label{initialstate}
\psi(x,y,0)=\frac{1}{\sqrt{\pi}}\mathrm{e}^{-(\mathrm{Im}\,x_0)^2-(\mathrm{Im}\,y_0)^2}\mathrm{e}^{-\frac{(x-x_0)^2}{2}-\frac{(y-y_0)^2}{2}}.
\end{equation}
The initial quantum-dot is a linear combination of its two internal states with equal amplitudes, and the two field modes are in Gaussian states corresponding to coherent field states; $|x_0/\sqrt{2}\rangle$ and $|y_0/\sqrt{2}\rangle$ respectively \cite{mandel}. Such initial states are readily prepared experimentally. We will further pick $y_0=0$ and $x_0=2\lambda$ such that the initial wave packet is approximately centered at the minima of the sombrero. Note that the initial average momentum is zero, and that $\Psi(x,y,0)$ is different from the adiabatic states (\ref{adstate2}). A consequence of this is that the wave packet evolution will not be restricted to a single APS. However, the upper adiabatic state is only marginally populated for our particular choice of initial state and the main phenomena studied here, the effects of the geometrical phase on the field properties, is indeed seen even though slight interference between the two adiabatic states occurs. Hence, we emphasize that the dynamics take place mainly on the lower adiabatic surface. Properties of the upper APS have been studied in Ref. \cite{uaps}.

We restrict the analysis to the cylindrically symmetric case, where the time evolved state 
\begin{equation}\label{timedepstate}
\Psi(x,y,\tau)=\frac{1}{\sqrt{2}}\left(\psi_e(x,y,\tau)\left[\begin{array}{c} 1 \\ 0\end{array}\right]+\psi_g(x,y,\tau)\left[\begin{array}{c} 0 \\ 1\end{array}\right]\right),
\end{equation}
will predominantly spread along the minima of the sombrero potential. As the wave packet broadens it will, after a certain time, start to self-interfere. We may estimate the characteristic time for this process by approximate the inherent spreading by free evolution along the minima of the sombrero potential to get
\begin{equation}
T_{in}\approx\sqrt{4\pi^2\lambda^2-1}\approx2\pi\lambda.
\end{equation}
Within this time, the wave packet width has expanded over a distance $2\pi\rho_{min}$.

From the full system state (\ref{timedepstate}), we can derive the reduced density operators for the separate constitutes
\begin{equation}
\rho_i(\tau)=\mathrm{Tr}_{j,k}\Big[\rho(\tau)\Big],
\end{equation}
where the subscripts represent, either the two modes $a$ and $b$ or the quantum-dot, and $\rho(\tau)=\Psi^*(x,y,\tau)\Psi(x,y,\tau)$. Using the reduced density operators we will especially study the photon statistics and the Husimi $Q$-distribution \cite{mandel,schleich}
\begin{equation}\label{quant}
\begin{array}{l}
\displaystyle{P_i(n)=\langle n|\rho_i(\tau)|n\rangle}, \\ \\
\displaystyle{\langle \hat{n}_i\rangle=\sum_nnP_i(n)}, \\ \\
\displaystyle{Q_i(\alpha)=\frac{1}{\pi}\langle\alpha|\rho_i(\tau)|\alpha\rangle}.
\end{array}
\end{equation}
Here, $|n\rangle$ is the $n$'th-photon Fock state, $|\alpha\rangle$ a coherent state with amplitude $\alpha$ and the subscript $i=a,b$ for the respective modes.

\subsection{Dynamics on the $T_{in}$ time scale}\label{subsec4a}
Discussed in Sec. \ref{sec2}, it is the term $V_{gauge}$ that gives rise to a geometric phase. To correctly describe the adiabatic evolution each term of $H_{JT}^{ad}$ in Eq.~(\ref{jtpolarad}) must be taken into account, and it is not enough to study dynamics upon the potentials $V_\pm^{ad}$. In this subsection we study the full dynamics using Hamiltonian (\ref{hamcav2}), and we hence go beyond any adiabatic approximation. However, in order to identify the effects of the geometrical phase we compare the results with the ones obtained by propagating the same initial state using the ``semi-adiabatic" Hamiltonian defined as
\begin{equation}\label{semiadh}
\tilde{H}_{JT}^{ad}=T+V_-^{ad},
\end{equation}
where $T$ and $V_-^{ad}$ are both given in Eq. (\ref{jtpolaradterms}). Accordingly, a wave packet evolving via the Hamiltonian $\tilde{H}_{JT}^{ad}$ around the origin will not accumulate any geometrical phase.  

\begin{figure}[ht]
\begin{center}
\includegraphics[width=8cm]{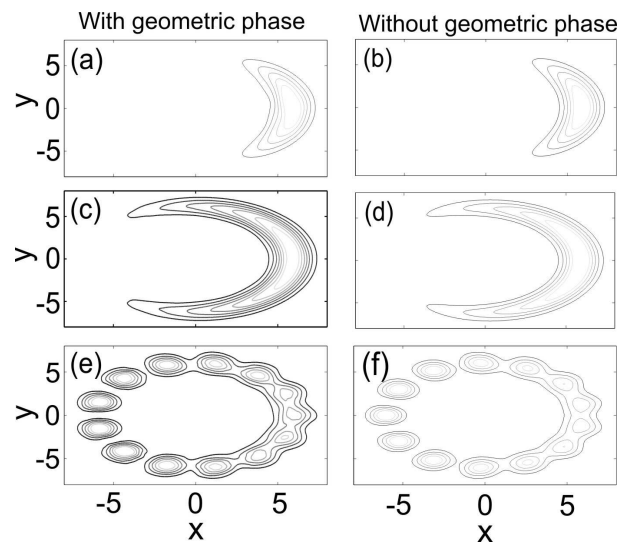}
\vspace{0cm} \caption{Snapshots of the wave packet distributions $|\Psi(x,y,\tau)|^2$ at times $\tau=T_{in}/4$ (a) and (b), $\tau=T_{in}/2$ (c) and (d), and $\tau=T_{in}$ (e) and (f) for the cases with (left) and without geometrical phase (right). In the last two plots, the difference between the interference structures is clearly visible. See Refs.~\cite{koppel,spinorbit2} for similar results. The dimensionless parameters are $\Omega=0.5$ and $\lambda=3$.   }
\label{fig3}
\end{center}
\end{figure}

The characteristic time scale $T_{in}$ determines how long it takes for the particular initial state (\ref{initialstate}) to inherently spread out across the CI and start to self-interfere. It is therefore a measure of the collapse time. The effect of the geometrical phase on the probability wave functions $|\Psi(x,y,\tau)|^2=|\psi_e(x,y,\tau)|^2/2+|\psi_g(x,y,\tau)|^2/2$ has been discussed in Refs. \cite{koppel,spinorbit2}. Initially we choose $x_0>0$ while $y_0=0$ such that interference of the evolved wave packet sets off at $-x_0$ where the two tails of the packet first join. Destructive and constructive interference cause nodes (vanishing probability distribution) and anti-nodes (non-vanishing probability distribution) in $|\Psi(x,y,t)|^2$, and the ring-shaped wave packet splits up in localized blobs. In the case of $\tilde{H}_{JT}^{ad}$, in which the geometric phase is zero, an anti-node builds up at $x=-x_0$, while for $\gamma_{JT}(\rho_{min})=-\pi$ (as is the case of zero detuning in the $E\!\times\!\varepsilon$ model) a node is formed at $x=-x_0$. The location of the corresponding node or anti-node depends on the value of $\gamma_{JT}(\rho_{min})$, and in all our examples $\Omega<\lambda$ such that $\gamma_{JT}(\rho_{min})\sim-\pi$ giving a node at $x\approx -x_0$. These features are visible in Fig.~\ref{fig3} showing the numerical results of the propagated distributions $|\Psi(x,y,\tau)|^2$ for three different times $\tau$. Full dynamics governed by Hamiltonian (\ref{hamcav2}), with $\phi=0$ and $\theta=\pi/2$ (cylindrically symmetric case) are shown in the left plots, while the right ones reproduce the results from propagation using the semi-adiabatic Hamiltonian (\ref{semiadh}). The effect of the geometrical phase becomes clear once the wave packet starts to self-interfere. The number of localized blobs depends on time $\tau$ and system parameters and then especially $\rho_{min}$. Note that very similar results where presented in Refs. \cite{koppel,spinorbit2}. However, the Hamiltonians used for the simulations in Refs. \cite{koppel,spinorbit2} are in general different from the one $\tilde{H}_{JT}^{ad}$ utalized here; a {\it single surface approximation} \cite{singlesurf,singlesurf2} is applied in most examples of Refs. \cite{koppel,spinorbit2} while here it is only considered for the non-geometrical phase case.  

\begin{figure}[ht]
\begin{center}
\includegraphics[width=6cm]{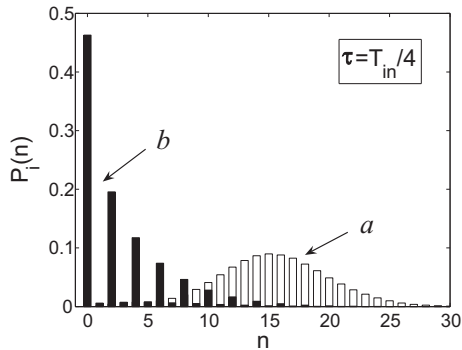}
\caption{The photon distributions $P_a(n)$ (white) and $P_b(n)$ (black) after a time $\tau=T_{in}/4$. The dimensionless parameters are the same as in Fig.~\ref{fig3}.  }
\label{fig4}
\end{center}
\end{figure}

As the initial wave packet starts to spread, a non-zero field will begin to build up in the vacuum $b$ mode, on the cost of decreasing field intensity of mode $a$. However, without the RWA, the total number of excitations is not conserved. In Fig.~\ref{fig4} we display the individual photon distributions $P_i(n)$ at a quarter of the interference time $T_{in}$. Already at this instant has the initially empty mode a non-zero field intensity, and its  photon distribution consists mostly of even number of photon states. This is a typical characteristic of Schr\"odinger cat states \cite{haroche}, and in the next subsection we will indeed show that such a state is created in the system at certain times. The small but non-zero population of odd photon numbers in the $b$ mode is caused by non-adiabaticity; for the semi-adiabatic Hamiltonian (\ref{semiadh}) the odd photon numbers are never populated for the given initial state (\ref{initialstate2}) with $y_0=0$.

\subsection{Dynamics beyond the $T_{in}$ time scale}\label{subsec4b}
Seen in the previous subsection, for an initial localized state mainly located at the minima of the sombrero potential and with zero average momentum, the time scale $T_{in}$ determines the collapse time; the time it takes for the localized wave packet to spread out over its accessible phase space region. Over longer periods, a revival structure in physical quantities is expected \cite{robinett}, where localized bumps are formed in phase space signalizing fractional or full revivals \cite{fracrev}. It has been pointed out however, that the collapse-revival characteristics are rather different in models where the RWA has been applied \cite{oldwine,dong}. Typically, in the RWA regime the various time scales become long. In this work we are outside such a regime, and we will in particular find that the revival time is given by a multiple of $\lambda T_{in}$ and that phase space evolution is significantly different for the two Hamiltonians (\ref{hamcav2}) and (\ref{semiadh}) due to the geometrical phase.   

From Fig.~\ref{fig3} we see that after a time $T_{in}$, the initial wave packet is spread throughout the minima of the sombrero potential and the self-interference causes nodes in the probability distribution. The number of localized bumps depends on $\rho_{min}$ (\ref{potmin}), but also on the time $\tau$; at first, when the self-interference sets off, the number of bumps increases to a maximum value and then the number begins to decrease and eventually form a single localized wave packet. Full revival occurs when a single localized bump is formed at the same position as the initial wave packet. To study the field dynamics we use the $Q$-function for the two modes $a$ and $b$ of Eq.~(\ref{quant}). We will present the two functions $Q_a$ and $Q_b$ in the same plots for brevity, but mark them with letters $a$ and $b$ respectively. At the initial time $\tau=0$, $Q_a$ and $Q_b$ are Gaussians centered at ${\bf \alpha}=(\sqrt{2}\lambda,0)$ and ${\bf \alpha}=(0,0)$ respectively. As time evolves, the $b$ mode builds up its intensity and the $Q$-function moves away from the origin, while $Q_a$ at first decreases its intensity by tending towards the origin. However, over longer time scales, $\tau>T_{in}$, a swapping of energy between the two modes will take place. This phenomenon has been discussed in our model, but only when the RWA has been imposed \cite{hamref2}. As our analysis concerns a regime far from the RWA one, this exchange of energy between the modes occurs at very different time scales than in Ref. \cite{hamref2}, similar to what was found for the inversion in the JC model \cite{oldwine}. Namely, the characteristic time scales in the parameter regimes of the RWA and without the RWA in the JC model can differ by orders of magnitude.

\begin{figure}[ht]
\begin{center}
\includegraphics[width=9cm]{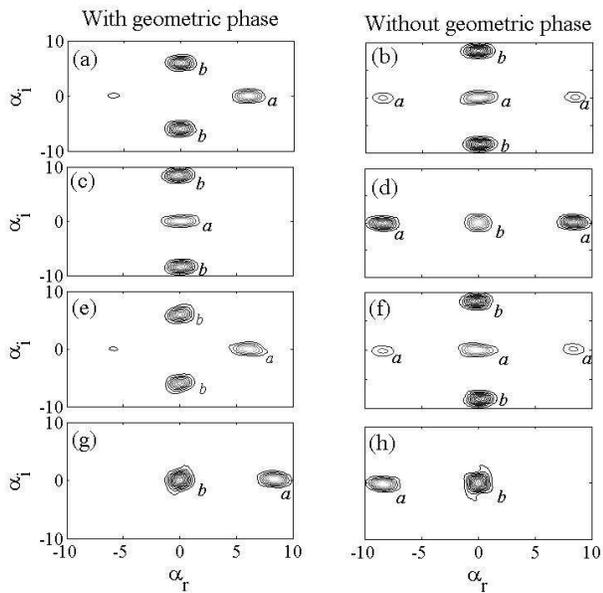}
\caption{Snapshots of the Husime $Q$-functions at times $\tau=\lambda T_{in}$ (a) and (b), $\tau=2\lambda T_{in}$ (c) and (d), $\tau=3\lambda T_{in}$ (e) and (f) and $\tau=4\lambda T_{in}$ (g) and (h). The left plots are the results with geometric phase, using Hamiltonian (\ref{hamcav2}), and the non-geometric phase results, obtained by the Hamiltonian in Eq.~(\ref{semiadh}), are displayed in the right figures. The $Q$-function of the $a$-mode is labeled by $a$ in the plots, while $b$ labels the second mode $Q$-function. The dimensionless parameters are $\lambda=6$ and $\Omega=0.5$. }
\label{fig5}
\end{center}
\end{figure}

From our numerical simulations we have found that localization of the phase space distributions comes about at multiples of time $T_{frac}=\lambda T_{in}$, which hence are the characteristic scales for fractional revivals \cite{robinett,fracrev}. The larger the radius $\rho_{min}$, the better resolved wave packet localizations. In Fig.~\ref{fig5} we display examples of the $Q$-functions $Q_a$ and $Q_b$ (indicated in the figures by $a$ and $b$) obtained either from the full system Hamiltonian (\ref{hamcav2}) (left) or from the semi-adiabatic Hamiltonian (\ref{semiadh}) (right). The times are here, $\tau=\lambda T_{in},\,2\lambda T_{in},\,3\lambda T_{in},\,4\lambda T_{in}$. A clear discrepancy is seen between the two models. For example, at $\tau=2\lambda T_{in}$ (c) and (d), mode $a$ in the left plot (with geometrical phase) is approximately in vacuum, while for the semi-adiabatic system (without geometrical phase), mode $b$ is roughly empty. At this instant, the non-empty mode is in a Schr\"odinger cat state. For $\tau=4\lambda T_{in}$ the full system has revived; the $Q$-functions have evolved into approximate replicas of their initial states. This is true up to an overall phase for the semi-adiabatic case, which is typical for a half-revival \cite{robinett}. From this figure we find the revival time for a wave packet encircling the CI in the $E\!\times\!\varepsilon$ model to be
\begin{equation}\label{revtim}
T_{rev}=4\lambda T_{in}\approx8\pi\lambda^2.
\end{equation}     
For the semi-adiabatic model, exact revivals (in terms of restoring also the overall phase) occur at twice this time. It should be pointed out that formula (\ref{revtim}) has been verified for a large set of different parameters.

\begin{figure}[ht]
\begin{center}
\includegraphics[width=6cm]{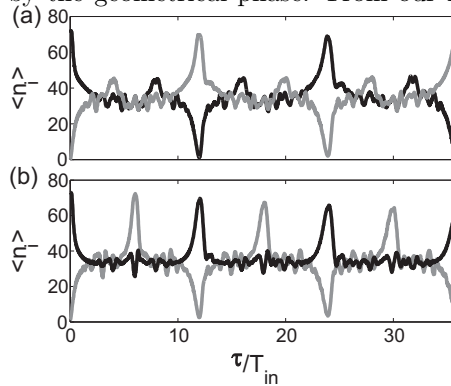}
\caption{The average photon numbers $\langle \hat{n}_i\rangle$ for both modes $a$ (black) and $b$ (gray) as a function of scaled time $\tau/T_{in}$ . The upper plot presents the results from using the full Hamiltonian (\ref{hamcav2}), which includes a geometrical phase, while the lower plot shows the results from using Hamiltonian (\ref{semiadh}). The effect of the geometrical phase is remarkably reflected in the two field intensities. The dimensionless parameters are the same as in Fig. \ref{fig5}. }
\label{fig6}
\end{center}
\end{figure}

Even though the phase space distribution of a cavity mode is in principle measureable \cite{tomo}, the field intensity is directly regained from the cavity output field using a photon-counter detector. Already Fig.~\ref{fig5} indicates that the average number of photons $\langle n_i\rangle$ differ considerably between the full model and the semi-adiabatic one. This is verified in Fig.~\ref{fig6} showing the time evolution of $\langle n_a\rangle$ and $\langle n_b\rangle$ for both models. Judging from the field intensities in this figure, the revival time of the semi-adiabatic model seems to be half the one of the full model, but here the $a$ mode is indeed not in a coherent state but in a Schr\"odinger cat. 

\section{Conclusions}\label{sec5}
In this paper we have shown how a system of a two-level ``particle" interacting with the fields of a bimodal cavity may fall in the category of JT models. By representing the model Hamiltonian in terms of field quadrature operators, rather than boson ladder operators, we identified its APSs and a CI. In this nomeclature, and in particular for the multi-particle analogue (Dicke model), the JT effect of cavity QED was identified with the normal-superradiant phase transition. The system studied here was described by the well known $E\!\times\!\varepsilon$ Hamiltonian. Knowledge from earlier research on this model, almost exclusively in molecular or chemical and condensed matter physics, has been applied on this cavity QED counterpart. Our main interest concerned the geometrical Berry phase reign from encircling the CI. The effect of the geometrical phase was studied by comparing physical quantities, such as the field phase space distributions and the field intensities, obtained from the evolution of either the $E\!\times\!\varepsilon$ Hamiltonian or the semi-adiabatic one in which no geometrical phase occurs. Clear distinctions between the two models were found when the system is let to evolve over longer time periods. Energy is swapped between the two field modes, and this exchange is highly affected by the geometrical phase. From our numerical results we could as well present analytical expressions for the collapse-revival times for a wave packet encircling the CI in the $E\!\times\!\varepsilon$ model. 

In addition, by introducing the notion of a wave packet evolving on two coupled potential surfaces, a deeper understanding of cavity QED models is obtained. This work, analyzing the geometrical phase, serves as an alternative viewpoint of the phenomenon in comparison to previous studies such as Refs. \cite{vedralham,berrycav}. It is indeed believed that the wave packet method used here will give even more thorough insight into cavity QED problems, or even trapped ion systems where related CI models are expect to occur \cite{dong}. We plan to study, using the current approach, the dynamics of the Dicke normal-superradiant phase transition. Another project underway is to investigate the `` molecular Aharanov-Bohm effect" and ``molecular gauge theory" in terms of cavity QED models.     

\begin{acknowledgements}
The author would like to thank Dr. \AA sa Larson for inspiring discussions.
\end{acknowledgements}


\begin{thebibliography}{999}

\bibitem{JTorig} H. A. Jahn and E. Teller, Proc. R. Soc. London Ser. A {\bf 161}, 220 (1937).

\bibitem{JTeff2} M. C. M. O'Brien and C. C. Chancey, Am. J. Phys. {\bf 61}, 688 (1993); I. Bersuker, {\it The Jahn-Teller Effect}, (Cambridge University Press, Cambridge, 2006).

\bibitem{ci} C. A. Mead, J. Chem. Phys. {\bf 78}, 807 (1983); D. R. Yarkony, Rev. Mod. Phys. {\bf 68}, 985 (1996); D. R. Yarkony, J. Phys. Chem. A {\bf 103}, 8579 (1999). 

\bibitem{cibook} M. Baer, {\it Beyond Born-Oppenheimer}, (Wiley, New York, 2006).

\bibitem{JTeff1} R. Englman, {\it The Jahn-Teller Effects in Molecules and Crystals}, (Wiley, New York, 1972). 

\bibitem{JTcrystal} C. A. Bates, Phys. Rep. {\bf 35}, 178 (1978).

\bibitem{longuet-higgens} H. C. Longuet-Higgins, U. \"Opik, M. H. L. Pryce, and R. A. Sack, Proc. R. Soc. London Ser. A {\bf 244}, 1 (1958). 

\bibitem{mt} C. A. Mead and D. G. Truhlar, J. Chem. Phys. {\bf 70}, 2284 (1979).

\bibitem{ab} Y. Aharanov and D. Bohm, Phys. Rev. {\bf 115}, 485 (1959).

\bibitem{berrymol1} A. Bohm, A. Mostafazadeh, H. Kuizumi, Q. Niu, and J. Zwanziger, {\it The Geometric Phase in Quantum Systems}, (Springer, Berlin, 2003).

\bibitem{berrymol2} C. A. Mead, Rev. Mod. Phys. {\bf 64}, 51 (1992). 

\bibitem{berry} M. V. Berry, Proc. R. Soc. London Ser. A {\bf 392}, 45
(1984).

\bibitem{faseffects} G. Delacr\'etaz, E. R. Grant, R. L. Whetten, L. W\"oste, and J. W. Zwanziger, Phys. Rev. Lett. {\bf 56}, 2598 (1986); F. S. Ham, {\it ibid.} {\bf 58}, 725 (1987); J. Sch\"on and H. K\"oppel, Chem. Phys. Lett. {\bf 231}, 55 (1994).

\bibitem{singlesurf} R. Baer, D. M. Charutz, R. Kosloff, and M. Baer, J. Chem. Phys. {\bf 105}, 9141 (1991).

\bibitem{koppel} J. Sch\"on and H. K\"oppel, J. Chem. Phys. {\bf 103}, 9292 (1995).

\bibitem{JTso} H. A. Jahn, Proc. R. Soc. London Ser. A {\bf 164} 117 (1938).

\bibitem{spinorbit2} J. Sch\"on and H. K\"oppel, J. Chem. Phys.
{\bf 1008}, 1503 (1998).

\bibitem{ent} A. P. Hines, C. M. Dawson, R. H. McKenzie, and G. J. Milburn, Phys. Rev. A {\bf 70}, 022303 (2004).

\bibitem{chaos} H. Yamasaki, Y. Natsume, A. Terai, and K. Nakamura, Phys. Rev. E {\bf 68}, 046201 (2003); E. Majernikova, and S. Shpyrko, Phys. Rev. A {\bf 73}, 066215 (2006).

\bibitem{magnetic2} G. Liberti, R. L. Zaffino, F. Piperno, and F.
Plastina, Phys. Rev. A {\bf 76}, 042332 (2007).

\bibitem{berrycav} I. Fuentes-Guridi, A. Carollo, S. Bose, and V. Vedral, Phys. Rev. Lett. {\bf 89}, 220404 (2002); S. Bose, A. Carollo, I. Fuentes-Guridi, M. F. Santos, and V. Vedral, J. Mod. Opt. {\bf 50}, 1175 (2003); A. Carollo, I. Fuentes-Guridi, M. F. Santos, and V. Vedral, Phys. Rev. Lett. {\bf 92}, 020402 (2004). 

\bibitem{vedralham} A. Carollo, M. F. Santos, and V. Vedral, Phys. Rev. A. {\bf 67}, 063804 (2003).

\bibitem{berrycav2} X. Z. Yuan and K. D. Zhu, Phys. Rev. B {\bf 74}, 073309 (2006);
S. Siddiqui, and J. Gea-Banacloche, Phys. Rev. A {\bf 74}, 052337 (2006).
 
\bibitem{oldwine} J. Larson, Physica. Scr. {\bf 76}, 146 (2007); J.
Larson, J. Phys: Conf. Ser. {\bf 99}, 012011 (2008).

\bibitem{eb} G. Levine and V. N. Muthukumar, Phys. Rev. B {\bf 69},
113203 (2004).

\bibitem{jc} E. T. Jaynes and F. W. Cummings, Proc. IEEE {\bf 51},
89 (1963); B. W. Shore and P. L. Knight, J. Mod. Opt. {\bf 40},
1195 (1993).

\bibitem{spinorbit1} C. A. Mead, Chem. Phys. {\bf 49}, 33 (1980); H. Koizumi, and S. Sugano, J. Chem. Phys. {\bf
102}, 4472 (1995); L. V. Poluyanov, S. Mishra, and W. Domcke, Mol.
Phys. {\bf 105}, 1471 (2007).

\bibitem{magnetic1} G. Bevilacqua, L. Martinelli, and G. P.
Parravicini, Phys. Rev. B {\bf 63}, 132403 (2001).

\bibitem{bobreak} H. K\"oppel, W. Domcke, and L. S. Cederbaum, Adv. Chem. Phys. {\bf 57}, 59 (1984); M. Baer, Phys. Rep. {\bf 358}, 75 (2002); G. A. Worth and L. S. Cederbaum, Ann. Rev. Phys. Chem. {\bf 55}, 127 (2004).

\bibitem{centcorr} A. S. Davydov, {\it Quantum Mechanics}, 2nd ed.
(Pergamon, Oxford, 1976).

\bibitem{vector} C. A. Mead, and D. G. Truhlar, J. Chem. Phys. {\bf
70}, 2284 (1979); C. A. Mead, Chem. Phys. {\bf 49}, 23 (1980).

\bibitem{haroche} S. Haroche and J. M. Raimond, {\it Exploring the
Quantum}, (Oxford University Press, Oxford, 2006).

\bibitem{qdot} A. Wallraf, D. I. Schuster, A. Blais, L. Frunzio, J. Majer, M. H.
Devoret, S. M. Girvin, and R. J. Schoelkopf, Nature {\bf 431}, 162
(2004); I. Chiorescu, P. Bertet, K. Semba, Y. Nakamura, C. J. P. M.
Harmans, and J. E. Mooil, Nature {\bf 431}, 159 (2004); N.
Hatakenaka, and S. Kurihara, Phys. Rev. A {\bf 54}, 1729 (1996);
 A. T. Sornborger, A. N. Cleland, and M. R. Geller,
Phys. Rev. A {\bf 70}, 052315 (2004).

\bibitem{bec} S. Slama, G. Krentz, S. Bux, C. Zimmermann, and P. W. Courteille,  Phys. Rev. A {\bf 75}, 063620 (2007); F. Brennecke, T. Donner, S. Ritter, T. Bourdel, M. K\"ohn, and T. Esslinger, Nature {\bf 450}, 268 (2007); P. Truetlein, D. Hunger, S. Camerer, T. W. H\"ansch, and J. Reichel, Phys. Rev. Lett. {\bf 99},  140403 (2007); Y. Colombe, T. Steinmetz, G. Dubois, F. Linke, D. Hunger, and J. Reichel, Nature {\bf 450}, 272 (2007); H. T. Ng, Phys. Rev. A {\bf 77}, 033617 (2008).

\bibitem{scully} M. O. Scully and M. S. Zubairy, {\it Quantum Optics},
( Cambridge University Press, Cambridge, 1997).

\bibitem{twomode} A. Messina, S. Maniscalco, and A. Napoli, J. Mod.
Opt. {\bf 50}, 1 (2003).

\bibitem{hamref} G. J. Papadopoulos, Phys. Rev. A {\bf 37}, 2482 (1988); C. Wildfeur and D. H. Schiller, Phys. Rev. A {\bf 67}, 053801 (2003).

\bibitem{hamref2} G. Benivegna and A. Messina, J. Mod. Opt. {\bf 41}, 907 (1994). 

\bibitem{dicke} R. H. Dicke, Phys. Rev. {\bf 93}, 99 (1954).

\bibitem{dicke2} Y. K. Wang and F. T. Hioe, Phys. Rev. A {\bf 7}, 831 (1973); K.
Hepp and E. H. Lieb, Ann. Phys. {\bf 76}, 360 (1973); {\it ibid.},
Phys. Rev. A {\bf 8}, 2517 (1973); N. Lambert, C. Emary, and T. Brandes, Phys. Rev. Lett. {\bf 92}, 073602 (2004).

\bibitem{a2} K. Rzazewski, K. Wodkiewicz, and W. Zakowicz, Phys.
Rev. Lett. {\bf 35}, 432 (1975); I. Bialynicki-Birula and K.
Rzazewski, Phys. Rev. A {\bf 19}, 301 (1979).

\bibitem{ramandicke} F. Dimer, B. Estienne, A. S. Parkins, and H. J. Carmichael
, Phys. Rev. A 75, 013804 (2007); S. Morris and A. S. Parkins, {\it ibid.} {\bf 77}, 043810 (2008).

\bibitem{jonasdicke} J. Larson, K. Rzazewski, and M. Lewenstein,
(unpublished).

\bibitem{adbose} M. Alexanian and S. K. Bose, Phys. Rev. A {\bf 52}, 2218
(1995).

\bibitem{cond} The detuning is independent as a control parameter, but it should be remembered that it can only be varied within the validity regime of the adiabatic elimination approximation.

\bibitem{photonmeas} M. J. Holland, D. F. Walls, and P. Zoller, Phys. Rev. Lett. {\bf 67}, 1716 (1991); D. I. Schuster, A. A. Houck, J. A. Schreier, A. Wallraff, J. M. Gambetta, A. Blais, L. Frunzio, J. Majer, B. Johnson, M. H. Devoret, S, M Girvin, R. J. Schoelkopf, Nature {\bf 445}, 515 (2007). 

\bibitem{squezmeas} H. M. Wiseman and G. J. Milburn, Phys. Rev. A {\bf 47}, 642 (1993); Q. A. Turchette, N. P. Georgiades, C. J. Hood, H. J. Kimble, and A. S. Parkins, Phys. Rev. A {\bf 58}, 4056 (1998). 

\bibitem{tomo} K. Vogel and H. Risken, Phys. Rev. A {\bf 40}, 2847 (1989); D. T. Smithey, M. Beck, M. G. Raymer, and F. Faridani, Phys. Rev. Lett. {\bf 70}, 1244 (1993); L. G. Lutterbach, and L. Davidovich, Phys. rev. Lett. {\bf 78}, 2547 (1997); 
P. Bertet, A. Auffeves, P. Maioli, S. Osnaghi, T. Meunier, M. Brune, J. M. Raimond, and S. Haroche, Phys. Rev. Lett. {\bf 89}, 200402 (2002); XuBo Zou, K. Pahlke, and W. Mathis, Phys. Rev. A {\bf 69}, 015802 (2004); A. A. Houck, D. I. Schuster, J. M. Gambetta, J. A. Schreier, B. R. Johnson, J. M. Chow, L. Frunzio, J. Majer, M. H. Devoret, S. M. Girvin, and R. J. Schoelkopf, Nature {\bf 449}, 328 (2007). S. Deleglise, I. Dotsenko, C. Sayin, J. Bernu, M. Brune, J. -M. Raimond, and S. Haroche, arXiv:08091064.

\bibitem{uaps} L. V. Poluyanov, S. Mishra, and W. Domcke. Chem. Phys. {\bf 332}, 243 (2007).

\bibitem{mandel} L. Mandel, and E. Wolf, {\it Optical Coherence and Quantum Optics}, (Cambridge University Press, Cambridge, 1995).

\bibitem{schleich} W. P. Schleich, {\it Quantum Optics in Phase Space} (Wiley, New York, 2001).

\bibitem{singlesurf2}S. Adhikari and  G. D. Billing, J. Chem. Phys. {\bf 111}, 40 (1999).

\bibitem{robinett} R. W. Robinett, Phys. Rep. {\bf 392}, 1 (2004).

\bibitem{fracrev} E. Romera and F. de los Santos, Phys. Rev. Lett. {\bf 99} 263601 (2007).

\bibitem{dong} D. Wang, T. Hansson, \AA. Larson, H. O. Karlsson, and J. Larson, Phys. Rev. A, {\bf 77}, 053808 (2008); H. Moya-Cessa, A. Vidiella-Barranco, J. A. Roversi, D. S. Freitas, and S. M. Dutra, {\it ibid.} {\bf 59}, 2518 (1999).




\end{thebibliography}
\end{document}